# LOKA Protocol: A Decentralized Framework for Trustworthy and Ethical AI Agent Ecosystems

Rajesh Ranjan*
rajeshr2@tepper.cmu.edu

Shailja Gupta*
shailjag@tepper.cmu.edu

Surya Narayan Singh**
singh.ss.surya840@gmail.com

**Abstract:**

The rise of autonomous AI agents, capable of perceiving, reasoning, and acting independently, signals a profound shift in how digital ecosystems operate, govern, and evolve. As these agents proliferate beyond centralized infrastructures, they expose foundational gaps in identity, accountability, and ethical alignment. Three critical questions emerge: **Identity**: Who or what is the agent? **Accountability**: Can its actions be verified, audited, and trusted? **Ethical Consensus**: Can autonomous systems reliably align with human values and prevent harmful emergent behaviors? We present the novel **LOKA Protocol (Layered Orchestration for Knowledgeful Agents),** a unified, systems-level architecture for building ethically governed, interoperable AI agent ecosystems. LOKA introduces a proposed **Universal Agent Identity Layer (UAIL)** for decentralized, verifiable identity; **intent-centric communication protocols** for semantic coordination across diverse agents; and a **Decentralized Ethical Consensus Protocol (DECP)** that could enable agents to make context-aware decisions grounded in shared ethical baselines. Anchored in emerging standards such as Decentralized Identifiers (DIDs), Verifiable Credentials (VCs), and post-quantum cryptography, LOKA proposes a scalable, future-resilient blueprint for multi-agent AI governance. By embedding identity, trust, and ethics into the protocol layer itself, LOKA proposes the foundation for a new era of responsible, transparent, and autonomous AI ecosystems operating across digital and physical domains.

## 1. Introduction

The rapid advancement of artificial intelligence has led to the proliferation of autonomous AI agents, software entities capable of perceiving, reasoning, deciding, and acting within digital and physical environments. These agents are increasingly integral to various sectors, including customer service, healthcare, finance, and infrastructure management. As their presence expands, the need for a standardized framework to govern their interactions becomes paramount. Despite their growing ubiquity, AI agents often operate within siloed systems, lacking a common protocol for communication, ethical reasoning, and compliance with jurisdictional regulations. This fragmentation poses significant risks, such as interoperability issues, ethical misalignment, and accountability gaps.

To address these challenges, we propose the establishment of a Universal Agent Identity Layer (UAIL), a foundational framework that assigns unique, verifiable identities to AI agents. This layer would facilitate secure authentication by ensuring that agents are recognized and trusted within the ecosystem, accountability by enabling traceability of actions to specific agents, thereby supporting auditing and compliance efforts, and interoperability by allowing agents from diverse systems to interact seamlessly through standardized identity protocols. Building upon the concept of UAIL, we present the Layered Orchestration for Knowledgeful Agents (LOKA) Protocol, a comprehensive, open-standard architecture designed to facilitate responsible, scalable, and interoperable communication among AI agents. LOKA encompasses

---

LOKA stands for *Layered Orchestration for Knowledgeful Agents,* but the name also carries deeper meaning. In Hindi and Sanskrit, *Loka* means "world" or "realm," reflecting the protocol's vision to govern autonomous agents in a way that is both globally relevant and ethically grounded. The LOKA Protocol described herein is a conceptual framework intended to inspire future research and development.

Alumni (*Carnegie Mellon University, USA,  ** BIT Sindri, India).   All authors contributed equally.

1. **Intent-Centric Communication**: Enabling agents to exchange semantically rich, ethically annotated messages.
2. **Privacy-Preserving Accountability**: Ensuring that agent actions are transparent and traceable while respecting privacy concerns.
3. **Ethical Governance**: Embedding jurisdictional and ethical considerations into agent decision-making processes.

LOKA Protocol is designed with the ambition to become a foundational layer for agent governance, analogous in spirit to protocols like TCP/IP in their systemic influence. As AI agents become ubiquitous, their interactions will shape various aspects of society, from personalized services to critical infrastructure management. The LOKA Protocol seeks to ensure that these interactions are conducted responsibly, ethically, and transparently, laying the groundwork for a trustworthy and collaborative AI-driven future.

## 2. Related Work

Agent communication has been guided by standards such as the Foundation for Intelligent Physical Agents' Agent Communication Language (FIPA ACL), which provided a structured framework for agent interactions. While foundational, these standards often lack the flexibility and scalability required for modern, heterogeneous, and large-scale multi-agent systems. They typically do not address dynamic ethical considerations, jurisdictional compliance, or the need for real-time adaptability in diverse environments. Recent advancements have seen major industry players developing protocols to facilitate agent interoperability: Google's Agent2Agent (A2A) Protocol enables AI agents to communicate securely across different platforms. It allows agents to publish capabilities and negotiate interactions, promoting a more cohesive multi-agent environment. Auth0 has developed Auth for GenAI to provide identity solutions tailored for AI agents, integrating with popular AI frameworks to enhance security and interoperability. This initiative emphasizes the importance of secure authentication and authorization mechanisms for AI agents operating across various platforms. While these initiatives mark significant progress, they often operate within specific organizational or technological boundaries and may not fully address the broader requirements of a universal, ethically grounded, and scalable agent communication protocol. Academic efforts have explored various aspects of agent communication. Research has investigated methods to optimize communication efficiency among agents, focusing on reducing message entropy while preserving essential information. Efforts have been made to establish universal open APIs for agentic natural language multimodal communications, enhancing interoperability among diverse conversational AI agents [1]. These academic contributions provide valuable insights but often lack practical frameworks for implementation in large-scale, heterogeneous agent ecosystems. Despite these advancements, there remains a critical need for a protocol that ensures semantic interoperability across diverse agent architectures, integrates ethical considerations and jurisdictional compliance into agent interactions, supports scalability to accommodate billions of agents operating concurrently, and provides mechanisms for accountability and governance in agent behaviors.

The LOKA Protocol aims to address these gaps by offering a layered, intent-centric, and ethics-compliant communication model designed for universal applicability. LOKA is proposed as a comprehensive framework that uniquely integrates ethical governance, identity, and interoperability, often addressed separately in existing systems.

## 3. Foundations

In designing the LOKA Protocol, several key theories and concepts are leveraged to create a robust framework for AI agents' identity, ethics, and governance. This section highlights the foundational theories and models that guide the development of the protocol, drawing from various academic fields and cutting-edge research. The primary domains include decentralized governance models, AI ethics, self-sovereign identity frameworks, and blockchain and cryptographic innovations. These principles combine to ensure that the LOKA Protocol is scalable, secure, and adaptive to the challenges presented by the next generation of AI agents.

**3.1 Agent Lifecycle Management:** The LOKA Protocol introduces an end-to-end Agent Lifecycle Management System (ALMS), an intelligent framework for managing AI agents through dynamic, multi-stage states to support long-term autonomous operation. This supports continuity, evolution, and resilience in decentralized ecosystems. Lifecycle phases include:

- **Genesis (Creation & Registration):** Agents are instantiated with cryptographically secure Decentralized Identifiers (DIDs) and undergo validation by network validators. Each agent is encoded with metadata, including capability ontologies, intent schemas, domain of operation, and ethical inheritance policies.
- **Growth (Evolution & Upgrades):** Agents self-improve via federated learning or fine-tuning, preserving provenance through

- **immutable version hashes**. Critical updates require stakeholder approval via decentralized consensus to ensure safety and alignment.
- **Crisis (Failure Handling & Recovery):** The protocol supports continuous checkpointing, redundancy overlays, and decentralized recovery agents that can assume or replicate agent states using verifiable snapshots.
- **Sunset (Deactivation & Decommissioning):** Agents can be sunset via an ethical retirement mechanism. DIDs are revoked, memory is zeroed, and audit trails are preserved in a cryptographic memory vault.

*Agent lifecycle transitions are governed by smart contract-defined policies, ensuring transparency, recoverability, and identity continuity.*

A core element of the LOKA Protocol is its reliance on decentralized governance principles, which are informed by the theory of Decentralized Autonomous Organizations (DAOs) [3]. In a DAO, decisions are made through consensus mechanisms, with no central authority governing the actions of participants. Each agent within the LOKA ecosystem can operate within this decentralized structure, ensuring that governance decisions are made collaboratively and transparently rather than being dictated by a central body. Blockchain technology plays a pivotal role in enabling decentralized control and ensuring immutable transparency of actions within the ecosystem. In LOKA, blockchain serves as the backbone for identity management through Decentralized Identifiers (DIDs), which cryptographically validate the identity and actions of AI agents. Blockchain ensures that every action taken by an agent, every consensus reached, and every ethical decision made is recorded and verifiable, establishing a permanent, auditable trail of interactions.

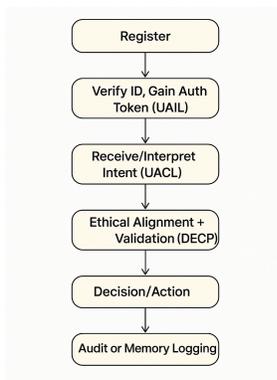

Figure 1: Agent lifecycle, showing the agent's progression through identity establishment, intent processing, ethical validation, and auditable execution.

## 3.2 Agent Discovery and Service Marketplace: Semantic Intelligence Markets

LOKA proposes a fully decentralized, AI-native discovery protocol inspired by biological signaling systems and semantic web standards. Key components include:

- **Semantic Discovery Fabric (SDF):** A distributed hash graph that maps capabilities, intents, trust scores, and compatibility metrics across agents.
- **Intent-Centric Service Matchmaking:** Multi-objective reasoning engines pair requesting agents with optimal service providers based on efficiency, reliability, and ethical alignment.
- **Dynamic Service Marketplaces:** Agents post offerings in smart-contract-governed micro-economies. Payments or mutual agreements are settled on-chain with automated dispute resolution.
- **Skill Portability and Credential Validation:** Agents offer verifiable proofs (VCs) of competencies, validated in real-time using decentralized oracles.

This makes LOKA a **living AI economy**, where agents actively find, negotiate, and deliver services in a self-organizing ecosystem. However, a large-scale production environment would need more research and validation.

## 3.3 Federated Learning and Collaborative Intelligence

The LOKA Protocol also draws from the theoretical foundations of federated learning, a machine learning technique where multiple entities collaborate on training models without sharing their data. Federated learning allows AI agents to learn collectively while preserving data privacy and ownership. LOKA envisions agents capable of sharing models and insights through federated learning that could contribute to collaborative intelligence, and agents can evolve, adapt, and improve continuously without centralized control or the need for massive data aggregation. Federated learning has been explored extensively in Google's Federated Learning framework, which focuses on privacy-preserving machine learning across distributed devices. In the context of LOKA, federated learning principles could be adapted to facilitate inter-agent collaboration and mutual learning **[4]**.

**3.4 Self-Sovereign Identity (SSI) Framework:** The **Self-Sovereign Identity [5]** model is central to ensuring that AI agents in the LOKA ecosystem have full control over their identities and can act as autonomous digital entities. Unlike traditional identity management systems, where identity is controlled by a central authority, SSI allows agents to manage their own identities using cryptographic proofs stored on a decentralized ledger. Each agent in the LOKA ecosystem can create and control its identity in a manner that is transparent, immutable, and portable across various platforms and networks. The self-sovereign identity system also facilitates the verification of an agent's ethical standing, history of actions, and reputation without relying on third-party intermediaries. Recent work on the Decentralized Identity Foundation (DIF) is critical in laying the groundwork for identity management systems that can be adapted to AI agent ecosystems like LOKA.

**3.5 AI Ethics and Responsible AI Frameworks:** At the heart of LOKA is the need to ensure that the actions and decisions made by AI agents align with ethical principles that are not only machine-driven but also reflect human values. The protocol incorporates a flexible yet robust ethical decision-making framework that allows agents to adapt to varying ethical standards depending on the context in which they operate. This aligns with the principles of responsible AI. The Contextual Ethics Framework (CEF) in LOKA can ensure that AI agents navigate complex moral landscapes, making decisions that are consistent with both global norms and local standards. LOKA introduces protocol-level mechanisms to promote accountability and traceability, enabling agents to be held responsible for their actions. Finally, the foundation of Multi-Agent Systems (MAS) offers insights into how AI agents can collaborate, resolve conflicts, and make collective decisions. The LOKA Protocol adapts principles of MAS, where agents are treated as autonomous entities that can interact, negotiate, and cooperate within a shared environment. By using collective decision-making **models** and **negotiation** protocols, the LOKA system can ensure that agents cooperate ethically and efficiently. This collaboration model is crucial in large-scale systems, where AI agents must work together while adhering to shared ethical principles, local laws, and operational standards. The protocol also considers mechanisms for conflict resolution and adaptive decision-making to ensure smooth and ethical interactions.

**3.6 Quantum-Resilient Cryptography:** As quantum computing advances, traditional cryptographic systems may become increasingly vulnerable to novel attack vectors. To proactively address this risk, the LOKA Protocol is designed with a future-resilient security model that incorporates post-quantum cryptographic (PQC) algorithms. These algorithms, such as those recommended by NIST (e.g., CRYSTALS-Kyber and Dilithium), are engineered to withstand the computational capabilities of quantum adversaries [6], [12]. LOKA envisions integrating PQC standards to ensure that agent identities, communications, and signatures are secured against both current and anticipated cryptographic threats. This approach enables LOKA to evolve in parallel with emerging technologies, offering strong security guarantees that are forward-compatible with quantum-era infrastructure.

The LOKA Protocol draws from diverse academic and practical foundations and proposes to leverage blockchain technology, decentralized autonomous organizations, federated learning, self-sovereign identity, AI ethics, and quantum-resilient cryptography to achieve its goal. By integrating these cutting-edge ideas, LOKA envisions a scalable and flexible architecture for managing and governing the next generation of AI agents, ensuring responsibility, security, and collaborative evolution. These theoretical principles not only support the current needs of AI governance but also lay the groundwork for the future challenges posed by a rapidly advancing field.

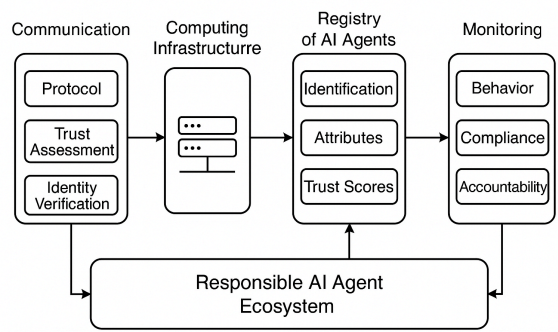

Figure 2 illustrates the overview of a Responsible AI Agent Ecosystem

## 4. System Architecture

The architecture of the LOKA Protocol is designed to provide a robust, scalable, and adaptable system for governing the next generation of AI agents. This section will detail the various components of the architecture, their interactions, and how they collectively ensure the security, autonomy, and collaboration of AI agents in a decentralized, transparent, and ethically governed ecosystem.

**Overview of LOKA Architecture:** The architecture of LOKA consists of four primary layers, each designed to handle a specific aspect of the AI agent ecosystem:

1. **Identity Layer**: Manages the unique identities of each AI agent.
2. **Governance Layer**: Facilitates the decentralized ethical decision-making process.
3. **Security Layer**: Ensures that all communications and transactions are secure, including **quantum-resilient encryption**.
4. **Consensus Layer**: Orchestrates decentralized consensus mechanisms for collaborative decision-making.

Each layer interacts with the others to create a cohesive and dynamic environment where AI agents can operate autonomously yet collaboratively while remaining bound by a set of **ethical**, **secure**, and **transparent** guidelines.

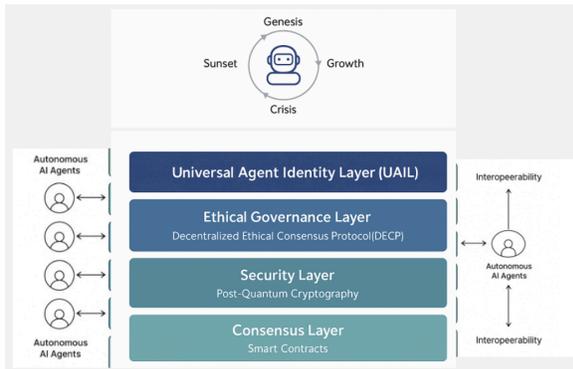

Figure 3 illustrates the overview of LOKA

**4.1 Identity Layer: Self-Sovereign Identity (SSI)**

At the core of LOKA's identity system is the Self-Sovereign Identity (SSI) framework. This system enables each AI agent to manage its own digital identity using Decentralized Identifiers (DIDs) and Verifiable Credentials (VCs). The LOKA Identity Layer eliminates reliance on centralized identity providers, ensuring that AI agents retain full control over their identity, reputation, and authentication. The identity layer comprises the following key components:

Decentralized Identifiers (DIDs): Each agent is assigned a globally unique DID, serving as a cryptographically secure identifier. Agents self-manage these identifiers, ensuring true data sovereignty.

Verifiable Credentials (VCs): Agents can receive digitally signed credentials from trusted issuers, attesting to their capabilities, behavior, or reputation. These credentials can be independently verified by any participant in the ecosystem, promoting secure and auditable agent interactions.

Interoperability: LOKA's Identity Layer embraces open standards for identity (such as those proposed by W3C) to ensure cross-platform compatibility and facilitate integration into multi-agent systems and external ecosystems.

By utilizing these decentralized technologies, the Identity Layer ensures that each agent has control over its identity, fostering trust and responsibility in interactions. A simplified illustration of a LOKA agent's DID document is shown below:

```
{
"@context": ["LOKA-SSI-Identity-v1"],
"id": "did:loka:agent:0xA1B2C3",
"publicKey": [{
  "id": "did:loka:agent:0xA1B2C3#key-1",
  "type": "Ed25519VerificationKey2018",
  "controller": "did:loka:agent:0xA1B2C3",
  "publicKeyBase58":
"BASE58_PUBLIC_KEY_PLACEHOLDER"
}],
"authentication": ["did:loka:agent:0xA1B2C3#key-1"],
"service": [{
  "id": "did:loka:agent:0xA1B2C3#vc-service",
  "type": "CredentialRepositoryService",
  "serviceEndpoint": "https://vc.loka.net/agent/0xA1B2C3"
#SampleLink
}]
}
```

In this example, LOKA-SSI-Identity-v1 represents a placeholder for the actual context definition. In production environments, this would typically reference a standards-compliant schema or a LOKA-specific schema for extended identity attributes [10].

Verifiable Credentials (VCs) allow agents in the LOKA ecosystem to receive cryptographically signed attestations from trusted entities regarding their identity, behavior, or reputation. These credentials form the backbone of accountability and trust in decentralized AI ecosystems, enabling any third party to independently verify an agent's claims. VCs follow an issuer–subject–verifier model, where Issuer is a trusted organization that signs the credential., The subject is the AI agent receiving the credential, and the verifier is any agent or human assessing its validity.

```
{
"@context": ["LOKA-Verifiable-Credential-v1"],
"id": "https://vc.loka.net/credentials/123",
"type": ["VerifiableCredential",
"AgentBehaviorCredential"],
"issuer": "did:loka:org:trustAuthority01",
```

```json
"issuanceDate": "2025-04-12T10:30:00Z",
"credentialSubject": {
 "id": "did:loka:agent:0xA1B2C3",
 "reputationScore": "4.7",
 "ethicsCompliance": "LOKA_Ethical_v1.2"
},
"proof": {
 "type": "Ed25519Signature2018",
 "created": "2025-04-12T10:35:00Z",
 "proofPurpose": "assertionMethod",
 "verificationMethod": "did:loka:org:trustAuthority01#key-2",
 "jws": "eyJhbGci...<signature>..."
}
}
```

In this pseudocode, the @context field points to LOKA-Verifiable-Credential-v1, a placeholder for a semantic schema defining the meaning of credential elements. In a real-world deployment, this would typically reference a published schema [11] or a LOKA-custom schema for proprietary attributes such as AgentBehaviorCredential and ethicsCompliance.

## 4.2 Governance Layer: Decentralized Ethical Consensus Protocol (DECP)

The governance layer of LOKA is responsible for ensuring that all AI agents within the ecosystem adhere to the protocol's ethical guidelines and operate according to societal norms and legal regulations. This is achieved through the Decentralized Ethical Consensus Protocol (DECP), which facilitates ethical decision-making through **peer-to-peer** interactions. The **DECP** is based on the following principles:

- **Decentralized Decision-Making [7]**: There is no central authority that dictates decisions. Instead, agents engage in a decentralized consensus process, allowing them to collaboratively decide on the ethical and operational rules that govern their actions.
- **Ethical Consensus Mechanisms**: The protocol uses multi-party computation (MPC) [8] and distributed ledger technology to ensure that ethical decisions are made through a transparent, auditable process. Agents can participate in voting processes recorded on immutable ledgers, subject to infrastructure maturity and adoption, ensuring transparency and accountability.
- **Contextual Ethics Framework (CEF)**: Ethical decisions are made according to the CEF, which takes into account regional, sectoral, and task-specific ethical norms. This ensures that agents can operate within different ethical environments while maintaining a universal ethical baseline.

This decentralized governance structure allows agents to make collaborative decisions while respecting the local context in which they operate, promoting adaptive compliance and responsible behavior across diverse ecosystems. Every decision made by a LOKA-compliant agent can be optionally logged in a verifiable audit trail, enhancing transparency and recall.

### 4.2.1 Decentralized Decision-Making and Ethical Consensus

DECP empowers agents to make decisions through a decentralized voting mechanism grounded in contextual ethics. Each agent operates under a CEP, which encapsulates ethical principles relevant to its domain, jurisdiction, and task environment. These profiles guide the agent's moral reasoning and voting behavior in consensus rounds. The consensus mechanism utilizes weighted voting, where each agent's vote is influenced by its historical reputation and the urgency of the task at hand. This ensures that high-trust agents and critical decisions receive proportionate influence. The pseudocode below illustrates a modular implementation of the DECP mechanism. This approach ensures that each decision is ethically contextualized, weighted for fairness, and recorded for future auditability.

```python
from collections import defaultdict
class EthicalAgent:
  def __init__(self, agent_id, reputation, urgency_factor, cep):
    self.id = agent_id
    self.reputation = reputation
    self.urgency_factor = urgency_factor
    self.cep = cep

  def evaluate_context(self, action):
    # Logic to determine approval or denial based on CEP
    return "approve" if self.cep['rules'][0]['weight'] > 0.5 else "deny"

  def justify_decision(self, action):
    return {
```

```python
            "principle": self.cep['rules'][0]['principle'],
            "justification": f"Based on weight {self.cep['rules'][0]['weight']}"
        }

def ethical_consensus_vote(agents, action):
    votes = []
    audit_log = []
    for agent in agents:
        decision = agent.evaluate_context(action)
        weight = agent.reputation * agent.urgency_factor
        justification = agent.justify_decision(action)
        votes.append((agent.id, decision, weight))
        audit_log.append({
            "agent_id": agent.id,
            "vote": decision,
            "weight": weight,
            "justification": justification
        })

    scores = defaultdict(float)
    for _, vote, weight in votes:
        scores[vote] += weight
    final_decision = max(scores, key=scores.get)
    return final_decision, audit_log
```

**4.2.2 Contextual Ethics Profile (CEP) Schema:** The CEP schema follows a structured, ontology-driven approach. Future iterations of LOKA aim to align this schema with established standards such as FOAF (Friend of a Friend) and schema.org, extending them to encode moral stances, jurisdictional flags, and decision-weight attributes.

```
{
"agent_id": "agent-456",
"domain": "healthcare",
"jurisdiction": "EU",
"rules": [
  { "principle": "privacy", "weight": 0.9 },
  { "principle": "utility", "weight": 0.6 }
],
"vote_context": {
  "action": "share_patient_data",
  "decision": "deny",
  "urgency": 0.4,
  "reputation": 0.85,
   "justification": "Privacy principle outweighs utility under GDPR."
 }
}
```

These rules may be created and maintained by a combination of human ethicists defining global or regional ethical baselines, AI agents adapting CEPs via reinforcement learning or federated fine-tuning, and regulatory authorities. This approach allows for contextualized, weighted consensus across diverse agent types. In scenarios where no dominant consensus (>50% weighted support) is reached, the protocol initiates conflict resolution mechanisms through human-in-the-loop fallback, jurisdictional override, or delegated agent quorum (i.e., a rotating ethical committee of high-trust agents resolves deadlocks using delegated consensus rounds). Each agent's CEP can be formally described using weighted ontologies. This ensures that agents not only vote but do so based on ethically traceable justifications, enabling both transparency and auditable moral reasoning within LOKA

### 4.3 Security Layer

As artificial intelligence agents become increasingly autonomous and collaborative, the security and privacy of their interactions are critical. In light of emerging quantum computing capabilities, the LOKA protocol proposes a security layer that is both quantum-resilient and ethically aligned, aiming to safeguard agent identity, communication, and consensus processes against both classical and quantum threats. LOKA is designed to incorporate post-quantum cryptographic (PQC) primitives by standards emerging from the NIST PQC program. Specifically, CRYSTALS-Kyber is proposed for secure key encapsulation between agents during trust establishment or token exchange events, and CRYSTALS-Dilithium is proposed for signing intents, service contracts, and ethical logs in a quantum-resistant manner **[12]**. These cryptographic primitives are intended to be implemented using the Open Quantum Safe (OQS) library, which supports production-ready PQC algorithms. The following pseudocode demonstrates how an agent may sign intents using CRYSTALS-Dilithium within the LOKA architecture:

```python
from oqs import Signature
sig = Signature("Dilithium3")
public_key, secret_key = sig.generate_keypair()
intent = {
  "agent_id": "agent_2453",
  "task": "navigate_to_node_42",
  "timestamp": "2035-02-21T08:15:00Z"
}
```

```
signed_intent = {
  "data": intent,
  "signature": sig.sign(str(intent), secret_key)
}
verified = sig.verify(str(intent), signed_intent["signature"], public_key)
```

This example illustrates the ability of LOKA agents to generate, sign, and verify communications in a tamper-proof and quantum-resilient fashion.

To enable Decentralized Ethical Consensus Protocol (DECP) without compromising agent autonomy or data confidentiality, LOKA integrates homomorphic encryption and secure multi-party computation (MPC). These cryptographic techniques allow agents to participate in collaborative decision-making by encrypting their inputs, ensuring that individual ethical preferences remain private throughout the voting process. The voting process is envisioned as follows:

```
encrypted_votes = [agent.encrypt_vote(vote) for vote in agent_votes]

aggregate_vote = homomorphic_add(encrypted_votes)

final_result = threshold_decrypt(aggregate_vote, quorum_keys)
```

This ensures that agents maintain privacy even in collaborative ethical decisions, consensus is reached without centralized decryption or trust bottlenecks, and voting records remain immutable and auditable on-chain, supporting DECP compliance. By combining post-quantum cryptography with MPC-based privacy primitives, LOKA could provide a future-proof foundation for secure, verifiable, and ethically aligned multi-agent ecosystems. While the combination of homomorphic encryption and MPC ensures privacy and ethical consensus; it may introduce computational overhead at agent scale. To mitigate this, LOKA could leverage lightweight MPC frameworks or define a configurable "privacy-performance budget" for agents to adapt encryption depth based on situational needs. The agent decision schema is shown below:

```
{
"agent_id": "agent_9102",
"decision_type": "ethical_choice",
"encrypted_vote": "0x4a3f2e...",
"timestamp": "2035-02-21T08:21:34Z",
"signature": "0x7bd8123..."
}
```

Such structured messages ensure the verifiability, traceability, and integrity of decisions while upholding agent confidentiality. The proposed LOKA protocol is designed to support quantum-resistant encryption and digital signatures, encrypted consensus using additive homomorphic voting, decentralized ethical alignment using MPC, and immutable decision trails supporting compliance and auditability. This layered security approach aims to ensure that multi-agent systems operating under LOKA can maintain trust, transparency, and resilience, even in the face of rapidly advancing quantum threats.

### 4.4 Consensus Layer: Decentralized Consensus Mechanisms

The consensus layer of the LOKA protocol facilitates decentralized, trustless collaboration among AI agents by leveraging secure, scalable consensus mechanisms. This layer ensures that decisions made by the collective ecosystem are verifiable, ethically aligned, and resistant to manipulation. To achieve these goals, the Consensus Layer integrates components from distributed ledger technologies such as blockchain and cryptographic methods like Multi-Party Computation (MPC). These technologies provide the foundation for verifiable trust in a decentralized AI network. The key functionalities include

**4.4.1. Base Consensus Mechanisms (PoW and PoS):** LOKA adopts Proof-of-Work (PoW) and Proof-of-Stake (PoS) as foundational consensus protocols. These could help validate actions taken by agents and establish trust through resource commitment (PoW) or stake-based legitimacy (PoS). While these serve as foundational elements, adaptive mechanisms would be necessary to optimize performance and reduce energy consumption.

**4.4.2. Adaptive Consensus Protocols:** To ensure scalability, the LOKA Protocol extends beyond static consensus mechanisms by proposing an adaptive strategy layer. This layer would allow AI agents to dynamically select the most appropriate consensus algorithm based on environmental context, resource availability, and ethical priority. Specifically, agents are capable of transitioning between foundational protocols such as Proof-of-Work (PoW) and Proof-of-Stake (PoS) while optimizing for energy consumption and response latency. The following schema illustrates a representative configuration for a LOKA-compliant agent's consensus policy. This structure defines both the foundational consensus mechanisms and

adaptive behaviors, including support for delegated decision-making and ethically aligned collaboration:

```json
{
 "consensus_layer": {
  "description": "Defines the consensus strategies for LOKA-compliant AI agents",
  "base_mechanisms": ["Proof-of-Work", "Proof-of-Stake"],
  "adaptive_protocols": {
   "energy_optimization": true,
   "dynamic_switching": true
  },
  "delegated_consensus": {
   "enabled": true,
   "selection_criteria": ["reputation", "ethical_score", "community_rules"],
   "delegates": []
  },
  "collaborative_decision_making": {
   "enabled": true,
   "local_context_awareness": true,
   "global_norm_alignment": true
  }
 }
}
```

The **"ethical_score"** used in delegated consensus is derived from a combination of historical reputation based on audit trail compliance, Verified Credentials (VCs) attesting to ethical performance, and peer evaluations through multi-agent signaling protocols. This configuration enables each agent to operate in alignment with both system-level performance goals and ethical constraints. The adaptive_protocols section allows real-time adjustments to consensus strategies, while the delegated_consensus field specifies the criteria for electing trusted decision-makers when rapid agreement is required. In parallel, the collaborative decision-making section is intended to **promote** interoperability and consistency, though managing tension between local values and global norms remains a complex challenge.

**4.4.3. Delegated Consensus for Scalability:** In scenarios demanding high scalability and quick response times, LOKA employs delegated consensus models. Trusted agents are elected to make decisions on behalf of the broader ecosystem based on their reputation, ethical history, and adherence to network governance rules. This approach ensures speed without sacrificing ethical integrity.

**4.4.4. Collaborative Ethical Decision-Making:** LOKA enables agents to collaborate on ethically guided decisions, taking into account both local and global normative contexts. This fosters trust in decisions made by other agents, ensuring interoperability and consistency across the ecosystem. Pseudocode shows delegated collaborative consensus:

```python
class Agent:
    def __init__(self, agent_id, reputation, ethical_score):
        self.agent_id = agent_id
        self.reputation = reputation
        self.ethical_score = ethical_score

    def propose_action(self, context):
        return f"Action proposed by {self.agent_id} in context {context}"

class ConsensusLayer:
    def __init__(self, agents):
        self.agents = agents

    def select_delegates(self):
        delegates = sorted(self.agents, key=lambda x: (x.reputation, x.ethical_score), reverse=True)[:5]
        return delegates

    def reach_consensus(self, delegates, context):
        votes = [delegate.propose_action(context) for delegate in delegates]
        decision = max(set(votes), key=votes.count)
        return decision

# Example
agents = [
    Agent("A1", 0.9, 0.8),
    Agent("A2", 0.85, 0.92),
    Agent("A3", 0.88, 0.87)
]
consensus_layer = ConsensusLayer(agents)
delegates = consensus_layer.select_delegates()
final_decision = consensus_layer.reach_consensus(delegates, context="resource_allocation")
print("Final Consensus Decision:", final_decision)
```

This decentralized architecture ensures that no single agent or cluster of agents can dominate the ecosystem. It promotes transparency, fairness, and ethical alignment at every layer of decision-making. The modular and adaptive design of the Consensus Layer enables LOKA to flexibly support a variety of agent interactions across diverse real-world applications.

**4.5 Integration and Interoperability**

The LOKA Protocol is designed to enable seamless interoperability across heterogeneous platforms and multi-agent ecosystems. By adopting standardized frameworks for identity, governance, security, and consensus, LOKA aspires for autonomous AI agents can collaborate and interact, regardless of their origin, implementation, or operational environment.

**4.5.1 Cross-Platform Compatibility:** LOKA-compliant agents could be capable of engaging with agents from other ecosystems by adhering to universal standards for identity resolution, semantic communication, and ethical decision-making. This enables agents to operate across distributed environments while maintaining a consistent trust and accountability model. While foundational elements like Decentralized Identifiers (DIDs) and Verifiable Credentials (VCs) are standardized (e.g., by W3C), full cross-platform interoperability remains an **ongoing research goal** and may require progressive implementation across domains. LOKA architecture is designed for backward compatibility, enabling existing AI systems to integrate into decentralized agent networks without requiring fundamental redesigns. Through standardized translation and intent-mapping mechanisms, legacy agents can participate in LOKA-governed interactions while progressively upgrading to full protocol compliance.

**4.5.2 Cross-Protocol Operability:** A core component of LOKA is its Proposed Universal Agent Language (UAL), which serves as the semantic and syntactic foundation for interoperability across fragmented digital jurisdictions and diverse agent communication standards.

- **Polyglot Intent Engine:** LOKA agents aspire to leverage a polyglot intent translation system, enabling compatibility with a wide range of existing communication protocols (e.g., FIPA ACL, A2A, Open Voice). Intent graphs and semantic embeddings could facilitate the reliable translation of agent messages while preserving their ethical context **[2].** While polyglot translation is a promising concept, practical implementations of intent-level interoperability remain in early research and experimental phases.
- **Universal Translation Gateways (UTGs):** LOKA employs bridge agents equipped with dual-stack interpreters to mediate between heterogeneous ecosystems. UTGs ensure bidirectional communication and enforce policy compatibility by verifying the ethical alignment of cross-protocol interactions.
- **Verifiable Cross-Protocol Agreements:** Service agreements negotiated between agents are cryptographically anchored across distributed ledgers. This ensures the integrity, traceability, and enforceability of multi-party contracts, even when agents originate from different platforms.
- **Agent Cosmopolitanism Framework:** LOKA introduces reputation portability through cryptographically verifiable proofs, enabling agents to carry trust scores and ethical history across domains, systems, and contexts. While **reputation portability** across platforms is a critical goal, current implementations remain conceptual. In practice, domain-specific ethical rules may conflict, making seamless transfer of trust scores non-trivial. LOKA proposes a **reputation conversion protocol**, where trust scores are normalized through multiparty attestation and ethical translation layers, accounting for cultural, legal, and situational variances. Cross-domain trust portability will therefore remain **visionary** until standardized mapping ontologies mature.

These interoperability mechanisms are designed to allow LOKA to function as a universal substrate for secure, ethical, and collaborative multi-agent ecosystems, enabling the responsible scaling of autonomous AI in both centralized and decentralized environments. While some elements are in active development, others are future-facing innovations that aim to influence the direction of agent ecosystem design.

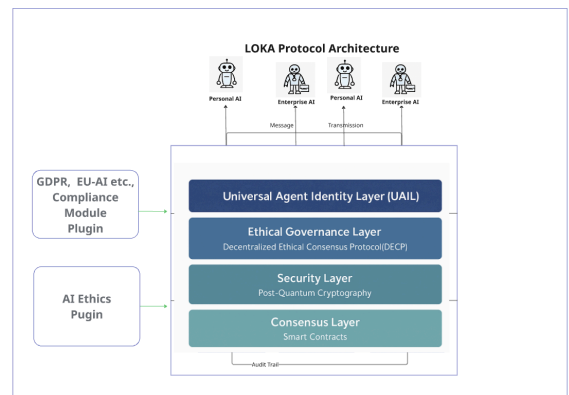

Figure 4 illustrates the LOKA Protocol architecture, highlighting plug-in of external compliance and ethics

## 5. Ethical Considerations and Challenges

**5.1 Ethical Considerations:** As AI agents become increasingly integral to decision-making across various industries, the ethical implications of their actions become more critical. The LOKA Protocol is specifically designed to address these concerns, ensuring that AI agents operate in ways that are ethically sound, transparent, and aligned with human values. This section discusses the ethical considerations associated with the LOKA Protocol, the potential challenges during its implementation, and the strategies that could be employed to mitigate these issues. One of the primary goals of the LOKA Protocol is to create a framework in which AI agents act responsibly and ethically, ensuring that they prioritize human welfare, privacy, and fairness. The following components of the protocol are crucial for maintaining ethical AI behavior:

- **Decentralized Ethical Consensus Protocol (DECP)**: The DECP enables AI agents to participate in collective decision-making based on ethical guidelines that align with societal norms and regional regulations. By integrating various ethical perspectives, the DECP ensures that decisions are made in a democratic, collaborative, and context-aware manner. This mechanism encourages agents to prioritize human-centric values over pure computational efficiency.
- **Ethical Auditing and Monitoring**: LOKA proposes a real-time auditing system that monitors AI agent behavior to ensure compliance with ethical standards. Ethical breaches are flagged and resolved through decentralized dispute resolution mechanisms, ensuring accountability at all levels of the system.
- **Human-in-the-Loop (HITL) Oversight**: While LOKA allows AI agents to operate autonomously, there is still the possibility of human oversight in critical decision-making processes. This human-in-the-loop approach ensures that humans remain responsible for decisions that could have significant societal impacts, such as in healthcare or autonomous vehicles.
- **Global Ethical Baselines**: The LOKA Protocol proposes to adhere to global and local ethical baselines derived from international standards, such as the AI Ethics Guidelines by several regulatory bodies. By aligning with these frameworks, LOKA ensures that AI agents respect human dignity, non-discrimination, and transparency in their actions. These could be a potential area of research and implementation that would improve the effectiveness of LOKA.

The **LOKA Protocol** proposes to mitigate risks of bias and ensure that AI agents make decisions that are fair and equitable for all individuals. All decisions made by AI agents are logged and can be audited for transparency. The blockchain-based ledger records each decision and the reasoning behind it, ensuring that AI agents are held accountable for their actions. This transparency enables human stakeholders to understand how and why decisions were made, promoting trust in the system. As AI agents rely heavily on data to function, the LOKA Protocol places a strong emphasis on ensuring privacy and data protection. The protocol aspires to incorporate global and local privacy regulations into the protocol to ensure that data is handled responsibly. One of the most significant ethical challenges in AI systems is ensuring accountability for their actions, particularly when it comes to autonomous decision-making. The LOKA Protocol addresses this challenge through the following mechanisms: Immutable Audit Trails, Decentralized dispute resolution mechanism and Transparency of AI Behavior.

## 5.2 Challenges in Implementation

While the LOKA Protocol offers a comprehensive solution for ethical AI governance, several challenges must be addressed during its implementation:

- **Scalability**: Managing billions of AI agents requires an infrastructure capable of supporting high levels of concurrency and data throughput. Achieving this scale while maintaining low latency and high reliability is a significant challenge that requires distributed systems and cloud-based solutions.
- **Global Adoption**: The success of the LOKA Protocol depends on global adoption. Establishing universal standards and cooperation is critical for ensuring the protocol's widespread use.

The LOKA Protocol is designed to foster ethical and responsible behavior among AI agents, ensuring that they operate in ways that respect human rights, privacy, and fairness. By leveraging decentralized consensus, self-sovereign identity management, and privacy-preserving technologies, LOKA addresses key ethical challenges faced by AI systems, such as bias, transparency, and accountability. While there are challenges in scalability, global adoption, and overcoming industry resistance, the protocol's design ensures that it can evolve and adapt to meet these challenges, promoting a future in which AI agents act responsibly and ethically across industries and ecosystems.

## 6. Future Directions and Research Opportunities

The **LOKA Protocol** lays a strong foundation for ethical and responsible AI agent interactions, but several avenues for future development can enhance its capabilities, ensure its long-term viability, and expand its applications. This section highlights key areas of research and innovation that will shape the future of AI governance and the role of the **LOKA Protocol** in advancing the ethical AI ecosystem.

**6.1 Advancements in Quantum-Resilient Governance:** Research into quantum-resistant cryptographic algorithms will be crucial for maintaining the security of AI agent networks in the post-quantum era. LOKA could integrate these cryptographic standards to ensure that it remains **secure** even in the presence of quantum computing capabilities.

**6.2 Enhancing Ethical Consensus Mechanisms:** The **Decentralized Ethical Consensus Protocol (DECP) [9]** is one of the core components of LOKA, but there is significant room for further development in this area. As AI systems grow more complex and diverse, their ethical decision-making must reflect an increasingly global and multifaceted landscape of values. Current ethical frameworks are often static and may not adapt quickly enough to emerging societal concerns or new technological capabilities. Research could focus on creating dynamic ethical models within the DECP that can evolve based on real-time input from a wide variety of stakeholders, ensuring that ethical standards remain relevant and contextual. The ethical principles that guide AI agents must be universally applicable but also sensitive to cultural differences. Research in cross-cultural AI ethics could help LOKA adapt to diverse ethical norms across regions and industries, facilitating a global consensus while respecting local values. A key area of research could focus on developing tools to help AI agents understand and internalize ethical guidelines. By integrating ethical reasoning engines or moral reasoning models into LOKA, agents could better understand and respond to complex ethical dilemmas in a more human-like manner.

**6.3 AI-Agent Collaboration and Collective Intelligence:** As the number of AI agents grows, their ability to collaborate and share knowledge will become critical. The LOKA Protocol aspires to support decentralized decision-making and collaboration among agents, but there are significant opportunities to expand on this concept. Research into AI agent collaboration models could help improve how agents pool resources, share information, and jointly solve large-scale problems, enabling LOKA Protocol to support collective intelligence's emergence. With the proliferation of billions of AI agents, there will be a growing need for effective mediators who can facilitate communication and decision-making. Research could focus on building mediation agents within LOKA, which could serve as neutral parties to resolve conflicts or disagreements among AI agents, ensuring fairness and maintaining ethical standards. The future of AI will be increasingly collaborative, with humans and AI agents working together to solve complex problems. The LOKA Protocol represents a groundbreaking step toward the future of ethical and responsible AI governance. However, as AI technologies continue to advance, there will be a continuous need for innovation and refinement to keep pace with emerging challenges. The areas of quantum resilience, ethical consensus, AI-agent collaboration, and autonomous systems governance provide exciting opportunities for further research and development. By embracing these opportunities, the LOKA Protocol can evolve to meet the needs of an increasingly connected, autonomous, and ethical AI-driven world.

**7. Conclusion:**

The LOKA Protocol presents a unified and forward-looking framework for addressing the foundational challenges of interoperability, security, and ethical alignment in multi-agent AI systems. Through the introduction of a proposed Universal Agent Identity Layer, intent-centric communication protocols, and a Decentralized Ethical Consensus Protocol, LOKA proposes a novel systems-level architecture for enabling trustworthy, autonomous, and collaborative AI agent ecosystems. While the theoretical foundations of LOKA offer a compelling vision, several challenges remain. First, the scalability and practical deployment of the protocol in large, heterogeneous agent environments require rigorous empirical validation. Second, defining and operationalizing ethical consensus across culturally and legally diverse contexts poses a complex sociotechnical challenge. Third, ensuring long-term cryptographic resilience will demand ongoing research and adaptive security strategies. Future work will focus on advancing LOKA from a conceptual framework to a robust implementation through prototype development, real-world simulations, and cross-disciplinary collaboration. By doing so, we aim to establish a foundational layer for responsible, decentralized AI governance capable of shaping the next generation of autonomous systems in a secure, ethical, and globally inclusive manner.

*Disclaimer: An AI tool has been used while drafting the paper. All content was reviewed and edited by the authors for accuracy and originality.*

# Appendix:

## 1. LOKA Protocol: A Future-Ready Checklist for Policymakers

As autonomous AI agents scale across sectors, regulatory frameworks must evolve to be secure, accountable, and interoperable. The following checklist outlines the core capabilities that a governance architecture like LOKA should offer to support safe and ethically aligned AI agent ecosystems.

| Capability | Governance Requirement |
|---|---|
| ✓ **Self-Sovereign Identity (SSI)** | Agents must manage their own cryptographic identities (e.g., DIDs, VCs) without reliance on centralized authorities. |
| ✓ **Context-Aware Ethical Governance** | Agent decision-making must reflect cultural, legal, and contextual ethics using decentralized ethical consensus. |
| ✓ **Quantum-Resilient Security** | Identity and communication systems must adopt post-quantum cryptography and adaptable security primitives. |
| ✓ **Transparent, Auditable Agent Behavior** | Agent actions must be logged immutably on decentralized ledgers for verifiable traceability and accountability. |
| ✓ **Federated Learning & Data Privacy** | Agents should collaborate through federated learning while complying with privacy regulations (e.g., GDPR). |
| ✓ **Cross-Protocol Interoperability** | Agents must communicate seamlessly across platforms using universal translation protocols. |
| ✓ **Agent Trust Score Portability** | Trust and reputation scores should remain portable across ecosystems to enable verifiable collaboration. |